\documentclass[a4paper,11pt]{amsart}
\begin{document}
\hyphenation{Schwarz-schild ge-ne-ra-ted gra-vi-ta-tio-nal
mea-ning bi-blio-gra-phy Per-ga-mon li-near}
\title[Wrong ``Id\'ees fixes'' in GR] {{\bf Wrong ``Id\'ees fixes'' in GR}}
\author{Angelo Loinger}
\date{}
\address{Dipartimento di Fisica, Universit\`a di Milano, Via
Celoria, 16 - 20133 Milano (Italy)}
\email{angelo.loinger@mi.infn.it}
%%\thanks{}

\begin{abstract}
Six wrong \emph{id\'ees fixes} concerning the black holes and the
gravitational waves.
\end{abstract}

\maketitle

\small \vskip0.10cm\par\hfill {``\emph{Id\'ee fixe}, n. Idea that
 \par\hfill dominates the mind, monomania.''
\par\hfill From ``The Concise Oxford Dictionary''.}

\normalsize

\vskip1.00cm
%\section{}
\noindent {\emph{\textbf{Id\'ee fixe}} \bf N$^{\circ}$1} \par
\vskip0.10cm
 A widespread ``Vulgate'' of general relativity (GR) claims that
if Schwarzschild  problem -- i.e. the problem of the Einsteinian
gravitational field generated by a point mass $M$ at rest -- is
solved using co-ordinate-free methods (such as orthonormal bases,
\emph{etc.}), the result is necessarily the  \emph{standard} form
of solution, erroneously named ``by Schwarzschild'', in reality
due to Hilbert, Droste, Weyl. \par This is a wrong  \emph{id\'ee
fixe}: as a matter of fact, the procedure of the ``Vulgate'' is
impaired by a vulgar logical fallacy, cf. e.g. R.M. Wald,
\emph{General Relativity} (The University of Chicago Press,
Chicago; 1984), p.121, eq.(6.15): only by writing
$r^{2}(\textrm{d}\vartheta^{2}+\sin^{2}\vartheta
\textrm{d}\varphi^{2})$ for the angular part, this author has
already chosen the standard HDW-form of solution. Remark that
Birkhoff theorem \cite{1} is not in question: indeed,
Schwarzschild's \emph{\textbf{original}} form of solution \cite{2}
is diffeomorphic to the \emph{exterior} part ($r>2m$; $m\equiv
GM/c^{2}$; $G$ is the gravitational constant) of the standard
form. The interior part ($0<r<2m$) of this form presents various
mathematical and physical defects \cite{3}; in particular, in the
complete manifold of the HDW-form, i.e. for $0<r<\infty$, it is
impossible to assign the \emph{time arrow} for every time
geodesic, according to rational criteria (private communication by
S.Antoci). On the contrary, this difficulty does not exist for the
\emph{original} Schwarzschild's form \cite{2}, which is regular in
the entire spacetime, with the only exception of the origin $r=0$
of the space co-ordinates, seat of the point mass $M$.

\vskip0.80cm
%\section{}
\noindent \noindent {\emph{\textbf{Id\'ee fixe}} \bf N$^{\circ}$2}
\par \vskip0.10cm
The undue privilege granted to standard HDW-form of  solution of
Schwarzschild problem and some physically senseless reflections on
its interior part ($0<r<2m$) have generated: \emph{i}) the fictive
notion of black hole (BH); \emph{ii}) the \emph{id\'ee fixe}
according to which the continued gravitational collapse of a
sufficiently massive celestial body must give origin to a BH. \par
Thus, \emph{Chandra Digest} of February 18th, 2004 tells us that
``Thanks to two orbiting X-ray observatories, astronomers have the
first strong evidence of a supermassive black hole ripping apart a
star and consuming a portion of it. The event, captured by NASA's
Chandra and ESA's XMM-Newton X-ray Observatories, had long been
predicted by theory, but never confirmed \ldots until now.'' \par
This is science fiction! \emph{In reality}, the observations
reveal only the action of the very strong \emph{\textbf{tidal
effects}} produced by an enormous mass concentrated in a
``punctual'' volume. The \emph{original} Schwarzschild's
$\textrm{d}s^{2}$ \cite{2}, for instance, interprets perfectly the
observational data. \par The continued gravitational collapse can
be described very generally in a simple and concise way. Remark,
first of all, that the necessary and sufficient condition that a
Riemann-Einstein spacetime admit the group of spatial rotations is
that its $\textrm{d}s^{2}$ be reducible to the following form,
where $r,\vartheta, \varphi$ are polar spherical co-ordinates
\cite{4}:

\begin{equation} \label{eq:one}
    \textrm{d}s^{2}=A_{1}(r,t)c^{2}\textrm{d}t^{2}-A_{2}(r,t)\textrm{d}r^{2}-A_{3}(r,t)\textrm{d}\omega^{2},
\end{equation}

with

\begin{equation} \label{eq:two}
    \textrm{d}\omega^{2}\equiv \textrm{d}\vartheta^{2}+\sin^{2}\vartheta
    \textrm{d}\varphi^{2}.
\end{equation}

 Let us consider a collapsing \emph{spherical} body of
mass $M$. As it is well known, the Einsteinian gravitational field
outside the body is \emph{time independent}, and can be described
consequently by the following $\textrm{d}s^{2}$ \cite{5}:

\begin{equation} \label{eq:three}
    \textrm{d}s^{2}=\left[1-\frac{2m}{f(r)}\right]c^{2}\textrm{d}t^{2}-\left[1-\frac{2m}{f(r)}\right]
    ^{-1}\left[\textrm{d}f(r)\right]^{2}-
    \left[ f(r)\right]^{2}\left(\textrm{d}\vartheta^{2}+\sin^{2}\vartheta
    \textrm{d}\varphi^{2}\right),
\end{equation}

where $f(r)$ is \emph{any} regular function of the radial
co-ordinate. If we put $f(r)\equiv r$, we have the standard
HDW-form; for $r\equiv
\left[r^{3}+\left(2m\right)^{3}\right]^{1/3}$ we obtain the
original Schwarzschild's form \cite{2}; \emph{etc}. \par The
\emph{physical} results are \emph{independent} of the choice of
the function $f(r)$ -- and this is true, in particular, for the
above communication of the \emph{Chandra Digest}.

\vskip0.80cm
%\section{}
\noindent \noindent {\emph{\textbf{Id\'ee fixe}} \bf N$^{\circ}$3}
\par \vskip0.10cm
It consists of the conviction that in the motions of
gravitationally interacting masses an essential role is played by
the gravitational waves. Now, it can be proved with the well known
method of Einstein-Infeld-Hoffmann (EIH) that in a first
approximation the above motions are Newtonian, then the successive
approximations yield some (relatively) small corrections. But
\emph{no} gravitational wave is present \cite{6}. In recent years
I have given several \emph{exact} (i.e. non-approximate) proofs of
the physical non-existence of gravitational waves in the motions
of the masses \cite{7}.

\vskip0.80cm
%\section{}
\noindent \noindent {\emph{\textbf{Id\'ee fixe}} \bf N$^{\circ}$4}
\par \nopagebreak \vskip0.10cm \nopagebreak
It is obvious that the notion of gravitational wave, as an object
endowed with a \emph{physical} reality, requires the existence of
a class of physically privileged reference frames. \par Fock tried
to prove quite generally that also in GR we have a class of this
kind: the set of the so-called \emph{harmonic} systems \cite{8}.
But Fock's arguments are not rigorous -- and on the other hand GR,
if properly understood, does not allow the existence of privileged
co-ordinate systems. Another vain attempt is due to Bondi \emph{et
alii} \cite{9}, who tried to prove the existence of a class of
privileged frames insofar as the gravitational waves solely are
concerned.

\vskip0.80cm
%\section{}
\noindent \noindent {\emph{\textbf{Id\'ee fixe}} \bf N$^{\circ}$5}
\par \vskip0.10cm
It regards the conceptual adequacy, for the treatment of the
gravitational waves, of the \emph{linearized} version of GR. Now,
in a beautiful article of 1944 Hermann Weyl proved the
mathematical and physical \emph{inadequacy} of the linear
approximation of GR, just under the above respect \cite{10}. It is
a pity that the astrophysical community ignore Weyl's
demonstration. \par It follows, in particular, from Weyl's paper
that the various computations regarding the behaviour of the
famous radiopulsar PSR1913+16 are fully destitute of a physical
value.

\vskip0.80cm
%\section{}
\noindent \noindent {\emph{\textbf{Id\'ee fixe}} \bf N$^{\circ}$6}
\par \vskip0.10cm
It consists of the settled belief that the gravitational wave
generates a certain gravitational field. Since this hypothetical
wave does not possess a true energy-momentum, but only a pseudo
(false) energy-momentum (a `mathematical fiction'', according to
Eddington) the above \emph{id\'ees fixe} is pure nonsense.

\normalsize \vskip0.5cm
%\section{}
\noindent {\bf Final remarks}
\par \vskip0.10cm
Many \emph{id\'ees fixes} regarding the BH's and the GW's are
contained, respectively, in the review articles by Celotti
\emph{et alii} \cite{11} and by Schutz \cite{12}.  \par In the
golden booklet by Albert Einstein \emph{The Meaning of Relativity}
(1955) there is no mention of BH's and GW's.\\

\end{document}